\documentclass{article}
\usepackage{amsmath}
\usepackage{graphics}
\usepackage{graphicx}
\usepackage{booktabs}
\usepackage{multirow}
\usepackage{caption}
\usepackage{placeins}
\usepackage{geometry}
\usepackage{adjustbox}
\usepackage{hyperref}
\usepackage{natbib}
\usepackage{titlesec}
\usepackage{float}

\titleformat{\section}
{\normalfont\Large\bfseries}
{\thesection.}
{1em}
{}

\titleformat{\subsection}
{\normalfont\large\bfseries}
{\thesubsection.}
{1em}
{}

\newcommand{\centeredsection}[1]{\begin{center}\bfseries #1\end{center}}

\graphicspath{ {./images/} }

\title{Amortized Neural Networks for Agent-Based Model Forecasting\footnote{The views expressed herein are solely those of the authors. The content and results of this research should not be considered or referred to in any publications as the Bank of Russia's official position, official policy, or decisions. Any errors in this document are the responsibility of the authors.}}

\author{Denis Koshelev\footnote{Bank of Russia}, Alexey Ponomarenko\footnote{Bank of Russia}, Sergei Seleznev\footnote{Bank of Russia}}
\date{}

\begin{document} 
\bibliographystyle{plainnat}
\maketitle
\section*{\centeredsection{Abstract}}
\vspace{-2em}
In this paper, we propose a new procedure for unconditional and conditional forecasting in agent-based models. The proposed algorithm is based on the application of amortized neural networks and consists of two steps. The first step simulates artificial datasets from the model. In the second step, a neural network is trained to predict the future values of the variables using the history of observations. The main advantage of the proposed algorithm is its speed. This is due to the fact that, after the training procedure, it can be used to yield predictions for almost any data without additional simulations or the re-estimation of the neural network.

\vspace{1em}
\textbf{JEL-classification: C11, C15, C32, C45, C53, C63.}

\textbf{Keywords: agent-based models, amortized simulation-based inference, Bayesian models, forecasting, neural networks.}

\newpage
\section{Introduction}
      Agent-based models (hereinafter, ABM) are gaining more and more popularity among economists both in the academic community and at economic institutions. As they are simulation based, they do not require theoretical solution and they allow the extension of various assumptions which are often incorporated into classical economic models, such as the homogeneity/representativeness of agents, their rationality, and the availability of full information (see \cite{12}, \cite{17}, and \cite{2}). However, their flexibility is also a disadvantage. Realistic ABMs often include hundreds (see \cite{8}) to millions of agents (see \cite{30}) and are represented as nonlinear state-space models that contain a number of latent variables which is proportional to the number of agents. Parameter estimation and forecasting in models of this size are computationally complex tasks and have not been fully studied in the literature.

      Many articles have been devoted to ABM parameter estimation (see the review in \cite{10}), and although this work is still in progress, the use of simulation-based inference (hereinafter, SBI, see \cite{7}) has allowed the achievement of some progress in this area. ABM forecasting remains mostly uncharted territory, however. In discussing future opportunities and challenges in ABM, \cite{2} describe the state of the field as follows:
      
      \textit{'Conventional macro-models are constructed in terms of aggregate variables that coincide with acquired data about the economy. This is convenient because the resulting data can be directly used to initialize these models. But ABMs are dynamical systems that model the world at the level of individual agents, such as households and firms. In order to run the model it is necessary to initialize the states of all of the individual agents in a way that is also compatible with aggregate measurements. Doing this properly requires complete micro-data on individuals, which is typically not available.}

      \textit{Absent such data it becomes necessary to invent plausible states for each individual agent so that the aggregate states of the model match the measured aggregate data. However the individual states in the model also need to compatible with each other and with the inherent dynamics of the model. If the states are not compatible in this sense, the model will generate transient behaviors that will result in poor forecasts. Given that model forecasts are never perfect, this is a recurrent problem - as time passes the forecasts of the model inevitably deviate from the measured aggregate data, and the initialization process must be repeated again and again. Finding good methods for doing this is an open problem that must be solved if we are to use ABMs for time series forecasting.'}

      In current practice, forecasting in ABMs is based either on initializing all states using microdata or their distributions and then simulating the data (see \cite{20}), as described by \cite{2}, or by using surrogate procedures (which do not require the initialization of the states of all agents) based on approximating joint distributions of the last observed data and the future values of the variables (see \cite{34}).\footnote{Although there are a number of articles on estimating parameters and hidden states based on particle filters (see \cite{25}), we have found none that use particle filters for prediction in ABM that are comparable in size to those used in practice.} The first method, as \cite{2} note, requires knowledge of a large amount of data that is not always available to researchers, while the second is computationally demanding for problems with more than two variables and has only approximate properties in terms of convergence to the posterior distributions of forecasts.

      In this paper, we propose a new method for unconditional and conditional forecasting in an ABM that allows predictions to be made without the initialization of all hidden states. The method is based on the idea of meta-learning (see \cite{13}) on simulated data\footnote{
      For examples of training predictive models on simulated data in the context of meta-learning, see \cite{18} and \cite{14}, which motivated the algorithm described in this paper.} and consists of two steps. In the first step, multiple artificial datasets are simulated, as in SBI. In the second step, an algorithm which predicts future values for a particular dataset from the history of observations is trained using a set of similar problems, as in meta-learning. In contrast to the previously proposed ones, our algorithm has the property of amortization, that is, once it is trained, it can be used for almost any data. Moreover, unlike forecasting on the basis of surrogate algorithms, such as by \cite{34}, the algorithm described below has good theoretical properties and should converge exactly to the posterior distribution of the forecast when the neural network is flexible enough.
      Despite the fact that the main interest for us is the application of the algorithm in the context of ABM, the procedure used is general enough and can be applied to a wide range of problems, so in Section 2, we describe the conditional and unconditional forecasting procedures in a general form, almost without relying on the specifics of ABM. Section 3 discusses the metrics that will be further used to evaluate the quality of the proposed algorithm. Section 4 is devoted to a description of the experimental results. In Section 5, we discuss several important issues concerning quality estimation and potential extensions of the algorithm. Section 6 presents the conclusion.

\section{Algorithm}
      We assume that the model is specified as a Bayesian state-space model with a prior distribution of model parameters $p(\theta)$, state equations $s_t \sim p(s|s_{t-1}, \theta)$, and observation equations $y_t \sim p(y|s_t, \theta)$,\footnote{In an ABM, this equation can have a stochastic form, as in the case of measurement errors, which are introduced into the model to mitigate small differences between the modeled and observed variables, or a deterministic form, as in the case where the model variables are strongly related to the observed ones.} where $\theta$ is the vector of model parameters, $s_t$ is the vector of the models hidden states at time $t$, and $y_t$ is the vector of observed variables at time $t$. It is additionally assumed that it is possible to simulate datasets consisting of the observed variables.

      Later in this section, we describe the unconditional forecasting procedure as well as its modification for the case of conditional forecasting.
      
\subsection{Unconditional forecasting}
      The main idea is that the prediction from the Bayesian state-space model at time $t$ can be represented as an estimation of states $y_{t+1},...,y_{t+h}$, which are not observable at that moment. To solve this problem, we modify the algorithm for estimating several characteristics of the marginal state distributions described by \cite{21}. The algorithm can be represented as a two-step procedure. In the first stage, a set of artificial datasets of different lengths $t$, each of which consists of the history of observed variables $x=\{y_1,...,y_t \}$ and the future values of these variables $y=\{y_{t+1},...,y_{t+h}\}$, is generated. In the second stage, a neural network is trained to predict $y$ from $x$. The formal description of the algorithm is presented below:

      \textbf{Algorithm 1. Amortized neural network algorithm for unconditional forecasting} ($N$ is the number of simulations, $T_{min}$ is the minimum length of the time series, $T_{max}$ is the maximum length of the time series, $h$ is the maximum forecasting horizon, and $L$ is the loss function)

1. Generation of artificial data
\[X=\{\},Y=\{\} \]

For $n=1,...,N$:

\hspace{1cm}1.a. Sample parameters from the prior distribution

\[\theta^n \sim p(\theta)\]

\hspace{1cm}1.b. Sample states conditionally on the parameters
\[s_0^n \sim p(s_0|\theta^n )\]
\[s_t^n \sim p(s|s_{t-1}^n, \theta^n ),\quad	t=1,..., T_{max}\]

\hspace{1cm}1.c. Sample observable variables conditionally on the parameters and states

\[y_t^n \sim p(y|s_t^n, \theta^n),	\quad	t=1,..., T_{max}\]

\hspace{1cm}1.d. For $\tau=T_{min},..., T_{max}-1$, create data
\[x^{n,\tau}=\{y_1^n,...,y_\tau^n \}\]
\[y^{n,\tau}=\{y_{\tau+1}^n,...,y_{\tau+min(h,T_{max}-\tau)}^n\}\footnote{A subset of the variables can be chosen as $y$ instead of the full set of observed variables. This may be especially useful when the dimension of the vector of observed variables is large.}\]

\hspace{1cm}and add them to the dataset 
\[X=X+\{x^{n,\tau} \} \]
\[Y=Y+\{y^{n,\tau}\} \]

2. Neural network training
\[\varphi^* = argmin_\varphi \sum_{x^i \in X, y^i \in Y} L(y^i, f_\varphi(x^i))\] 

      In this paper, we take the cross-entropy with a diagonal normal distribution as a loss function, which allows us to correctly estimate the mean and standard deviation of the forecasts, however, as noted by \cite{21}, any M-estimator (see Chapter 5 in \cite{33}) can be used. The architecture of the neural network is similar to the architecture for the DSGE model of \cite{21} and consists of convolutional, recurrent, and fully connected layers.\footnote{Four types of one-dimensional convolutional layers with 16 filters each and lengths of 3, 5, 7, and 9, respectively, a two-layer GRU block with a dimension of 64, and three fully connected layers with an intermediate dimension of 100 and ReLU activation.} The three key differences are 1) the use of one-directional recurrent blocks to avoid 'looking ahead', 2) the dimensionality of the final outputs, which in our case is equal to the dimensionality of $y$, and 3) the addition of skip connection between the inputs and outputs (see \cite{19}).

      The proposed algorithm has the property of amortization, i.e., after training, the neural network can be used for any datasets without running computationally complex procedures such as particle filter (see \cite{15}). Instead, the data are fed into the neural network as inputs, and the desired characteristics of the forecast distribution (the mean and standard deviation in our case) are obtained as outputs within hundredths or tenths of a second.

\subsection{Conditional forecasting}
      The main difference between conditional and unconditional forecasting is the presence of a scenario. To account for the presence of a scenario in Algorithm 1, we modify feature matrix $x$ by adding scenario variables $z$ to it, assuming that they can be expressed in terms of the observed variables. In this case, the prediction algorithm is written as:

      \textbf{Algorithm 2. Amortized neural network algorithm for conditional forecasting} ($N$ is the number of simulations, $T_{min}$ is the minimum length of the time series, $T_{max}$ is the maximum length of the time series, $h$ is the maximum forecasting horizon, and $L$ is the loss function)

1. Generation of artificial data
\[X=\{\},Y=\{\} \]

For $n=1,..., N:$

\hspace{1cm}1.a. Sample parameters from the prior distribution
\[\theta^n \sim p(\theta) \]

\hspace{1cm}1.b. Sample states conditionally on the parameters
\[s_0^n \sim p(s_0|\theta^n )\]
\[s_t^n \sim p(s|s_{t-1}^n, \theta^n ),\quad	t=1,..., T_{max}\]

\hspace{1cm}1.c. Sample observable variables conditionally on the parameters and states
\[y_t^n \sim p(y|s_t^n, \theta^n),	\quad	t=1,..., T_{max}\]

\hspace{1cm}1.d. For $\tau=T_{min},..., T_{max}-1$, create data
\[x^{n,\tau}=\{[y_1^n,z_1^n(y_1^n, y_2^n,...,...,y_{1+min(h, T_{max}-1)}^n)],...,[y_\tau^n,z_\tau^n(y_1^n,..., y_\tau^n, y_{\tau+ 1}^n,...,y_{\tau+min(h, T_{max}-\tau)}^n)]\}\]
\[y^{n,\tau}=\{y_{\tau+1}^n,...,y_{\tau+min(h,T_{max}-\tau)}^n\} \]

\hspace{1cm}and add them to the dataset
\[X=X+\{x^{n,\tau} \} \]
\[Y=Y+\{y^{n,\tau}\} \]

2. Neural network training
\[\varphi^* = argmin_\varphi \sum_{x^i \in X, y^i \in Y} L(y^i, f_\varphi(x^i))\] 

      The architecture of the neural network for Algorithm 2 is similar to the architecture of the neural network for Algorithm 1, except that scenario variables are added as inputs at the level of the recurrent layer. Note that the presence of scenarios formulated in terms of the observed variables allows the input of data at the level of the recurrent or convolutional layers, because it is necessary to know the scenarios in the past periods for the correct training of a neural network with such an architecture. In order to incorporate scenarios that potentially include unobserved variables (for example, to calculate impulse responses that depend on the history of observations) in the neural network architecture, they must be fed into the input of the neural network in layers after which knowledge of the scenarios in the other time periods is not required. For the architecture proposed in Section 2.1, this can be done in the fully connected layers.

In spite of the fact that this neural network architecture is more general in terms of the number of tasks to be solved, through additional experiments (see Appendix C), we find that such networks are often much more difficult to train and require a more flexible architecture in the fully connected layers. Therefore, we recommend using Algorithm 2 for models in which the scenario is based on the observable variables only.

\section {Performance metrics}
      The algorithms proposed in the last section should converge exactly to the mean and standard deviation of the posterior distribution of forecasts with the number of simulations tending to infinity and with a sufficiently flexible neural network. In practice, however, the number of observations and the size of the neural network are large, but nevertheless finite, and the loss function contains many local optima. All this can lead to the trained neural networks yielding results different from the posterior of the forecasts,\footnote {See the discussion of the quality of various simulation algorithms, including amortized neural networks, in the context of finding posterior distributions of parameters presented by \cite{23}} so checking the quality of the resulting approximations is an important step. In this section, we describe two metrics that we use to check the quality of neural network training.

Of course, the best way to answer the question about the quality of the approximation of the characteristics of the posterior forecast distribution is to compare it with the characteristics of the posterior forecast distribution itself. Unfortunately, for most simulation models, and in particular for ABM, which is the focus of this paper, constructing an exact posterior distribution of forecasts is impossible, and approximations of it based on MCMC algorithms (see \cite{1}) or sequential Monte Carlo algorithms (SMC; see \cite{4}) require computationally complex algorithms with a particle filter (see \cite{15}) or variations of it.

\textbf {Forecast error standardization.} In research on probabilistic time series forecasting,\footnote{See \cite{5}} the forecasts are often tested by interval calibration or probability integral transforms. These methods cannot be applied directly to a case in which only the mean and the standard deviation of the forecast are estimated, since the probability distribution of the forecasts is not fully specified. However, it can be noted that, similarly to the probability integral transforms, which must match the normal distribution and have zero autocorrelation, the standardized forecast errors (with the mean removed and divided by the standard deviation) should have the same properties (zero mean and unit standard deviation), other than the form of distribution,\footnote{The distribution need not be normal.} for a well-trained model. Thus, we look at the mean and the standard deviation of the standardized forecast errors and the mean and the standard deviation of the product of standardized errors separated by $k$ periods to test the quality. The latter, in fact, is equivalent to testing for autocorrelation, but it does not require the adjustment of the asymptotic distribution due to the finite length of the time series. Thus, there is no worry that the distribution of the correlation estimates will have a mean and standard deviation different from zero and one.

      There are two points to note about this quality metric. First, the results of the test described above should be considered one stage of the verification of the quality of the model. Not passing it should serve as a signal of problems with the quality of the neural network. The reverse situation, in which the test is passed, is not a guarantee that the model works well. For example, if the conditional forecasting model is poorly trained and does not take the presence of a scenario into account in any way, but produces only an unconditional forecast, it will pass the test. Second, we cannot apply the standard formal hypothesis testing about the mean and standard deviation, since the drift of the neural network coefficients at non-zero learning rates (see \cite{26}) makes a comparable contribution to the distribution of the mean and standard deviation estimates.\footnote{It is probable that the estimates could be improved by applying ensembles of models based on several runs of the training procedure, or by using averaging at different iterations within the same training procedure, but this issue is beyond the scope of this paper.} Therefore, below, we look at these quantities without formal hypothesis testing.

\textbf{Comparison with the benchmark model.} As mentioned above, passing the test on standardized forecast errors is only an indirect confirmation of the quality of the neural network, since, among other things, it can be passed by models that do not take all relevant information into account. To see how well the neural network takes historical information into account, we estimate the lower bound of forecast quality on a test dataset.

      Such an estimate can be made with a benchmark model. Note that after optimization, the neural network should have the smallest mean square forecast error (hereinafter, MSFE) for each variable and for each horizon by construction of the loss function. Moreover, the mean log predictive scores (hereinafter, LPS) should be the largest in the class of normal distributions. These two facts allow us to conclude that, no matter which forecasting model we build on the test data, it should not outperform the neural network in terms of MSFE and LPS.

      The benchmark model should be chosen based on a balance of flexibility and training time. On the one hand, the more flexible the model, the tighter the lower bound will be. On the other hand, estimating the lower bound should take adequate time. A representative test dataset often contains thousands or tens of thousands of time series sets, so it is necessary to train the model hundreds of thousands of times (the number of test sets multiplied by the number of periods in which forecasts are made) to test one forecast horizon for one variable.

      In this paper, vector autoregression (VAR) with ridge regularization is chosen as a benchmark model. It is computationally easy enough to estimate and allows the calculation of both conditional and unconditional predictions in adequate time. The VAR model in the experiments is put in as comfortable a setting as possible, making it more difficult to pass the test. We start predicting only from the 101st period to give the VAR model more information to train the coefficients (re-estimated recurrently on an expanding window) and also optimize the number of lags and the regularization parameter on the test data over the grid.

\section{Results of the experiments}
      In this section, we show how the proposed algorithm works when making forecasts in the ABM. However, before moving on to the ABM, where the comparison with MCMC is time consuming, we show in Section 4.1 how our procedure and quality metrics behave using the toy example of a Bayesian AR(1) regression. Then, in Sections 4.2 and 4.3, we show the results for conditional and unconditional forecasts in the ABM.

\subsection{Proof of concept}
      We use the Bayesian AR(1) model (see Appendix B.1 for details) 	to test the proposed algorithm. We generate 1,010,000 time series (10,000 are taken as the test dataset) with a length of 200 to estimate the model. We choose $T_{min}=50$ because we find a slight degradation in quality compared to MCMC for models with $T_{min}$ and $T_{max}$ differing by an order of magnitude. At the same time, time-series lengths from 50 to 200 correspond to quarterly data from 12.5 to 50 years, which is in line with the series lengths used by macroeconomists in practice. The final training dataset for each of the forecasting horizons is approximately 150,000,000 examples (${\sim}N(T_{max}-T_{min} )$) and 1,500,000 for the test. The model is trained with the ADAM algorithm (see \cite{22}) 500,000 iterations with a batch size of 100 and learning rate $\varepsilon_n$:

\begin{equation*}
\varepsilon_n = 
 \begin{cases}
   10^{-4}, if\quad n<3\times10^5 \\
   10^{-5}, if\quad n \geq 3 \times10^5
 \end{cases}
\end{equation*}
The training takes about 5 hours on an NVIDIA GeForce RTX 2070 GPU for a neural network implemented in PyTorch (see \cite{29}).

      Figure 1 (see Appendix A) presents examples of the forecasts of the neural network and predictions using the MCMC algorithm on randomly generated data (not used in training) with sample lengths of 50, 100, 150, and 200 on horizons from 1 to 12. It can be seen from the figure that the forecasts are well approximated by the neural network in all cases: autocorrelation is close to zero (upper graph), intermediate autocorrelation (two middle graphs) and high persistence (lower graph).

      Figures 2 and 3 show how the errors and the products of the forecast errors are distributed for different forecast horizons on the test data. The means and standard deviations are also shown in the titles of the figures. The neural network approximating the posterior distribution of the forecasts of the Bayesian AR(1) model passes the standardized error test, since for all distributions, the mean and standard deviation are close to 0 and 1. Note that the error distributions are close to normal, but as mentioned above, this is not a necessary condition for our verification\footnote{As we have found, this is not the case for unconditional prediction in the ABM (graphs are not presented for the sake of space).} in contrast to the probability integral transforms.

      Table 1 compares the MSFE and LPS neural network errors with respect to the AR(1) model, which is a data generation process for each time series. It can be observed that the neural network statistically significantly outperforms the AR(1) model on all forecast horizons. This may seem counterintuitive, but there is nothing strange in this, since the Bayesian model knows a little more information. In particular, it knows information about how the parameters are distributed over the test datasets in addition to the data generation process.

      To summarize, we can say that the neural network copes well with our toy example. The characteristics of the posterior distributions of the forecasts almost coincide with the characteristics of the forecasts based on the MCMC algorithm, which is the main measure of the quality of training. Moreover, this neural network easily passes the indirect quality tests, which are the main tests in our work with the ABM, as expected.
      
\subsection{Unconditional forecasts}
      To demonstrate the properties of the proposed algorithm for the ABM, we build a simplified version of the model of \cite{8}, which contains 50 C-firms, 500 consumers, the public sector, and a bank (see Appendix B.2). In all examples for unconditional and conditional forecasts, we generate 510,000 datasets (10,000 of which are the test data) of length 300, 100 of which are removed as a burn-in period. This is the minimum number of simulations to prevent overfitting in a model with four observable variables (see below) without regularization as we have seen from the experiments. Of course, regularization procedures such as early stopping, L1/L2-regularization, or dropout\footnote{See \cite{31}} can be used to reduce the number of simulations, but the selection of the correct regularization and its hyperparameters lies outside the scope of this paper.

In the case of the ABM, contrast to AR(1), the data simulations take most of the time compared to the training of the neural network. We vectorize our model to speed up the simulation procedure. This allows us to generate 10,000 simulations simultaneously in a single notebook in Python. We run 5 Python notebooks in parallel, which allows us to generate 50,000 simulations in about 3 hours on a CPU (Intel(R) Core (TM) i7-8750H CPU @ 2.20GHz, 16GB RAM). As a result, the final dataset is generated in 30 hours. The neural network training for each of the three sets of variables described below takes about 5 hours and is run with similar hyperparameters to the Bayesian AR(1) model.

      We estimate the model with two, three, and four observed variables:
      \[X_2=\{10 log(1+P), 10U\}\]
      \[X_3=\{10 log(1+P), 10U, 10 log(1+C) \}\]
      \[X_4=\{10 log(1+P), 10U, 10 log(1+C),log(1+L) \}\]
where $X_2$, $X_3$, and $X_4$ are datasets of two, three, and four variables, respectively, $P$ is the average price level of the C-firms, $U$ is the unemployment rate, $C$ is the consumption level, and $L$ is the volume of loans. The price, consumption, and credit variables are transformed using a transformation that is close to a log transformation,\footnote{Log transformation is usually applied to these variables in macroeconomic forecasting.} with a modification that allows us to train the model without additional filtering for degenerate cases (when the values in the data are zero). The data are also scaled so that the standard deviations of the variables in the individual datasets are of the order of 1 on average.

      Table 2 presents the means and standard deviations of the forecast error distributions for models trained on datasets of two, three, and four variables. It can be seen that, in all cases, the means are close to 0 and that the standard deviations are close to 1, which means that the trained neural networks pass the first part of the test proposed in Section 3. Deviations of a few hundredths are caused by a small oscillation of the coefficients at learning rates that do not tend to zero, as is mentioned in the description of the test. When calculating the standard deviations of the product of the standardized forecast errors, we truncate the sample by removing points that are greater than 30 in absolute value (about 100 of the 1,500,000 test points). This is done because the standard deviations of the product of the errors are the fourth moments of the error distribution, and the presence of outliers can significantly affect the estimates. These outliers are associated with very extreme single simulations for unemployment and loans. Examples of such simulations are presented in Figure 4. Excepting such cases, Table 3 shows that the characteristics are close to ideal and differ by no more than a few hundredths.

      The neural network approximating the ABM demonstrates good predictive properties compared to the benchmark VAR model. Table 4 presents a comparison in terms of MSFE and LPS. It can be seen that, for all models and forecast horizons, the neural network is not (statistically significantly) worse than the VAR model. Moreover, it almost always (in 212 of 216 cases) outperforms the benchmark at the 1\% significance level. Together with the results of the previous test, this allows us to hope for a reasonably good approximation of the posterior of the ABM forecasts. Examples of forecasts on randomly generated data for the ABM with four variables are presented in Figure 5.
      
\subsection {Conditional forecasts}
      The model of Section 4.2 is used to illustrate the properties of neural networks in an amortizing conditional forecast task in the ABM. We train a neural network to predict consumption for 12 periods, assuming that unemployment is known over the forecast horizon.\footnote{The variables are transformed in the same way as in Section 4.2.} These variables are strongly correlated in our model (the average absolute correlation is greater than 0.5), so unemployment is quite informative with respect to consumption, and the forecast should be very different from the unconditional forecast. This allows us to clearly show how Algorithm 2 works.

Figure 6 presents an example of running a trained neural network on the same random data for consumption and unemployment as in Figure 5. It demonstrates that the results of the conditional forecast are much more accurate and narrower than those of the unconditional forecasts from the model with four variables. This serves as indirect evidence that the neural network is well-trained. This is also signalled by the quality metrics proposed in Section 3. Figures 7 and 8 and the values of the mean and standard deviation of the distributions (which are close to 0 and 1) for the forecast errors and their products show that the neural network passes the test on standardized forecast errors.

      As before, we estimate a VAR as the benchmark model for the second test from Section 3. The conditional forecasts from the VAR are constructed using Kalman smoothing (see \cite{9}). Table 5 shows that the neural network outperforms the VAR model on all forecast horizons. Like the results described in the previous paragraph, this is also indirect evidence of the quality of the neural network.

\section{Discussion}
      In this section, we discuss issues that we consider important in the context of the future application and development of the algorithms described. In particular, these are questions that are related to 1) the estimation of the performance of the algorithms based on indirect metrics, 2) scenario forecasting tasks which take into account scenarios built on unobserved variables, 3) the use of an algorithm on microdata, and 4) working with a fixed computational budget on simulations.

\textbf{Use of indirect quality metrics.} The results of the previous section allow us to conclude that the proposed amortized forecasting procedure based on neural networks demonstrates adequate results. Nevertheless, the quality estimates for both the conditional and unconditional forecasts are based on indirect metrics. Although this cuts off inadequate results, it cannot serve as a 100\% guarantee of a good approximation of the posterior distribution of forecasts. The benchmark model serves only as a lower bound on the quality of the forecast and may not be tight enough. Verification based on standardized forecast errors has problems similar to the validation of Bayesian model parameters (see \cite{6}; \cite{32}), in the sense that this test can be passed even by models whose distribution characteristics are far from those of the posterior distribution.\footnote{For example, the prior distribution can pass this test.} Moreover, in the case of non-zero learning rates, the latter procedure is not formally strong due to the fact that it is not based on statistical tests. As mentioned in Section 3, the errors introduced by the oscillation of the coefficients are comparable to the statistical error or even exceed it. The use of indirect metrics creates the risk that the predictions of underfitted models (which differ from the posterior ones) will be accepted as true predictions from the ABM, which may affect decisions made based on these predictions.\footnote{Note that such risks are associated not only with our algorithm, but are also applicable to many other approximate forecasting methods in structural models.} We therefore consider the improvement of forecast quality testing procedures as one of the most important directions for the further development of forecasting with ABMs. Among other things, we see great potential for testing amortized algorithms in formalizing tests on standardized forecast errors and building tests for the dependence of forecast errors on previous data.

\textbf{Scenario forecasting based on unobservable variables.} We also believe, despite the fact that we fail to achieve convergence to the optimum of the algorithm for scenario analysis, which depends not only on the observable variables, but also on unobservable ones (see Appendix C), that future research will find neural network architectures and learning algorithms that are more successful and help solve a wider range of problems than only unconditional and conditional forecasting.

\textbf{Application of the algorithm to microdata.} In addition to estimation on aggregate data, when working with ABMs in practice, it is often necessary to work with models that are fitted to historical microdata or their distributions (see \cite{20}). Algorithms 1 and 2 described in Section 3 can also be used for such problems by adding appropriate variables to the set of observables. These are the microdata themselves for the case of microdata,\footnote{Modifications to the structure of the neural network that perform aggregation operations in the first layers or that take the structure of the data into account more specifically, such as the relationship graph structure (see \cite{11}), may be needed to reduce the time and improve the quality.} and sufficient statistics (if any) or a small set of characteristics describing the properties of the distribution in the case of distributions.\footnote{See, for example, \cite{3} for a representation of the dynamics of distributions through a set of characteristics in the context of functional VAR models.}

\textbf{Fixed budget for simulations.} Simulation time in AMBs can be large, and a single simulation can take several minutes. This can render the time required to generate the necessary amount of artificial data prohibitive. There are many multi-round methods in the SBI literature that work under such constraints (see \cite{28}, \cite{24}, \cite{16}). We suggest using the idea of the rectangular truncation of the prior distribution of parameters from \cite{27}. It can be used to extend our approach by focusing on simulations in regions in which the parameters are more or less consistent with the observed data. Despite the fact that, like other algorithms, this approach does not have the property of amortization, it is nevertheless locally amortized within the truncated regions. Explicit amortization boundaries make it possible to estimate local properties using the quality metrics described above. Moreover, as for other algorithms, local amortization in the parameter space allows us to hope that when new data (new points in the time dimension) arrive, the old simulations can be reused, since the posterior distribution of the parameters usually does not change much when several new points are added.

\section {Conclusion}
      In this paper, we describe a forecasting algorithm for ABMs based on amortized neural networks, which allows for the almost instantaneous estimation of the characteristics of the posterior distribution of forecasts after pre-training. The experimental results for Bayesian AR(1) and verification based on indirect quality metrics for ABM show that the algorithm demonstrates good properties. We hope that our work will serve as a starting point for the development of forecasting procedures in ABMs based on neural networks, and also help to make forecasting in ABMs a routine procedure in the future.

\newpage
\nocite{*}
\bibliography{lib}

\newpage
\section*{Appendix A. Figures and tables}
\begin{figure}[h]
  \centering
  \vspace{-15pt}
  \textbf{Figure 1. Comparison of neural network and MCMC algorithm forecasts for Bayesian AR(1) model}\par\medskip

\vspace{-5pt}
  \scalebox{0.95}{\includegraphics{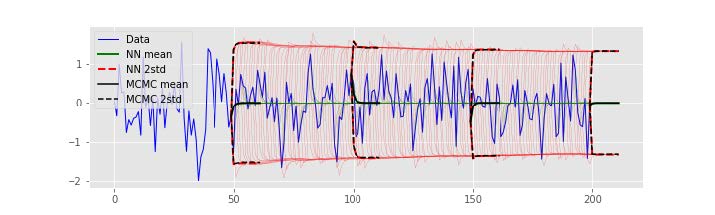}}  
\vspace{-7pt}
  \scalebox{0.95}{\includegraphics{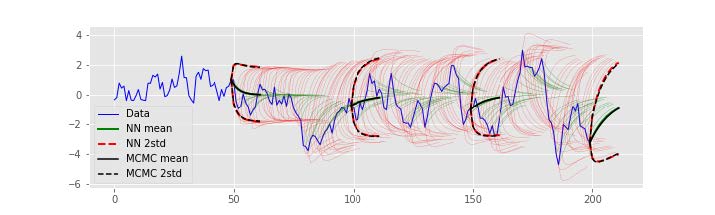}}
\vspace{-7pt}
  \scalebox{0.95}{\includegraphics{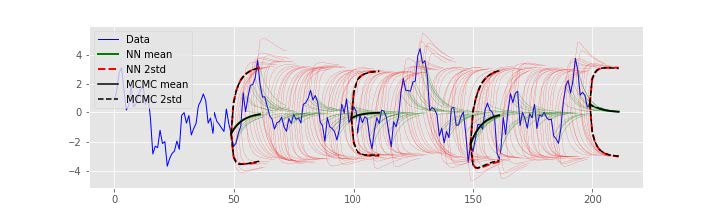}}
\vspace{-30pt}
  \scalebox{0.95}{\includegraphics{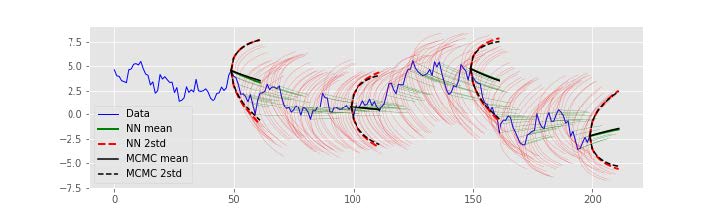}}
\end{figure}

\newpage
\begin{figure}[h]
  \centering
  \textbf{Figure 2. Distributions of standardized forecast errors for Bayesian AR(1) model}\par\medskip
  \scalebox{1}{\includegraphics{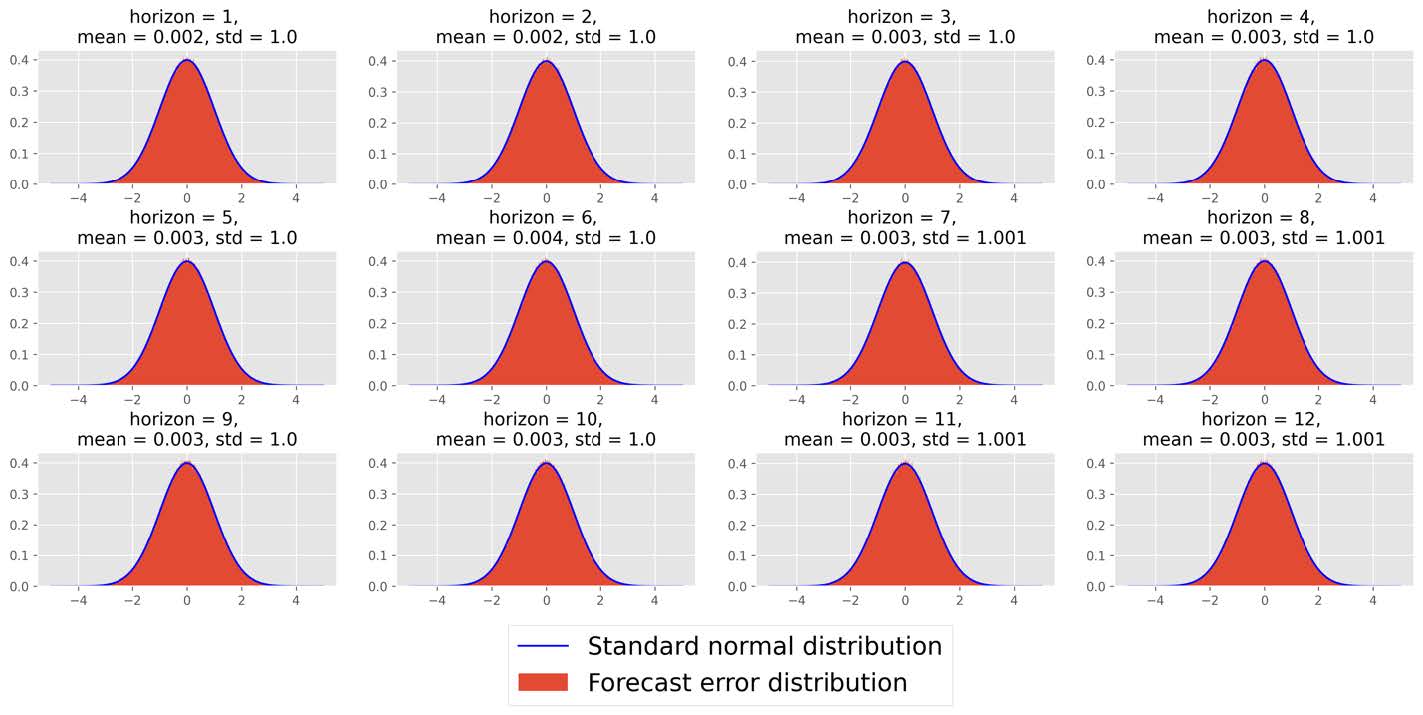}}
\end{figure}

\begin{figure}[h]
  \centering
  \textbf{Figure 3. Distributions of products of standardized forecast errors for Bayesian AR(1) model}\par\medskip
  \scalebox{1}{\includegraphics{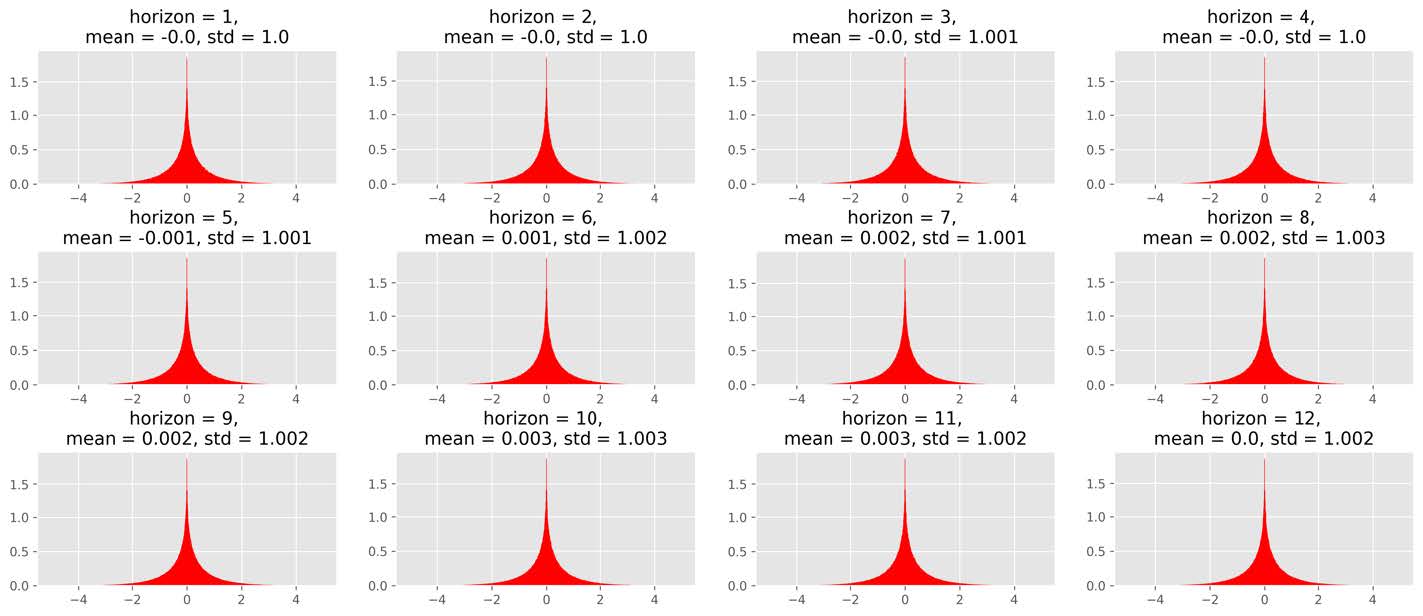}}
\end{figure}
\FloatBarrier
\newpage
\newgeometry{left=2.5cm, right=2cm, top=1.2cm, bottom=2.5cm}
\begin{table}[t]
\centering
\captionsetup{font=bf, justification=centering}
\caption{\textbf{Comparison of forecasting properties of AR(1) model and neural network approximating Bayesian AR(1) model (MSFE is the ratio of the mean squared errors of the AR(1) and the neural network; LPS is the difference in the mean log predictive scores of the AR(1) and the neural network)}}

\begin{tabular}{c|cc}

\textbf{Forecasting horizon} & \textbf{MSFE} & \textbf{LPS} \\
\midrule
1 & 1.009*** & -0.005*** \\
2 & 1.015*** & -0.008*** \\
3 & 1.02*** & -0.009*** \\
4 & 1.025*** & -0.011*** \\
5 & 1.029*** & -0.012*** \\
6 & 1.033*** & -0.013*** \\
7 & 1.038*** & -0.014*** \\
8 & 1.041*** & -0.015*** \\
9 & 1.045*** & -0.016*** \\
10 & 1.049*** & -0.016*** \\
11 & 1.052*** & -0.017*** \\
12 & 1.055*** & -0.018*** \\

\end{tabular}
\caption*{\textnormal{Significance level: *(10\%), **(5\%), ***(1\%)}}
\end{table}

\FloatBarrier
\begin{figure}[h]
  \centering
  \textbf{Figure 4. Examples of extreme simulations of unemployment and loans in ABM (graphs show different simulations)}
  \scalebox{1}{\includegraphics{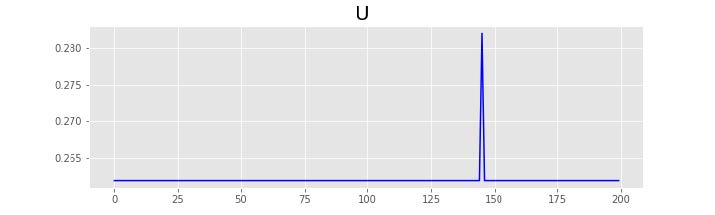}}
  \scalebox{1}{\includegraphics{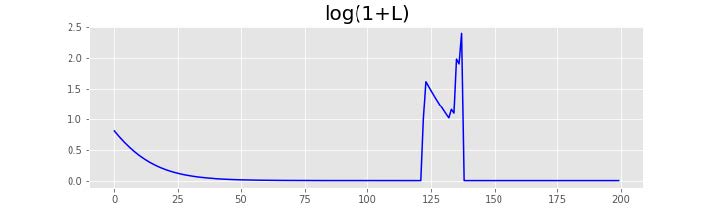}}
\end{figure} 
\restoregeometry
\FloatBarrier
\newpage

\newgeometry{left=2.5cm, right=2cm, top=2.5cm, bottom=2.5cm}
\begin{table}[h]
\centering
\captionsetup{font=bf, justification=centering}
\caption{\textbf{Verification of distributions of standardized forecast errors for unconditional forecast in ABM}}
\begin{tabular}{c|cccc|cccccc}
\multicolumn{1}{c|}{} & \multicolumn{4}{c|}{\textbf{2 variables}} & \multicolumn{6}{c}{\textbf{3 variables}} \\
\multicolumn{1}{c|}{\textbf{Forecasting horizon}} & \multicolumn{2}{c}{\textbf{CPI}} & \multicolumn{2}{c|}{\textbf{Unemployment}} & \multicolumn{2}{c}{\textbf{CPI}} & \multicolumn{2}{c}{\textbf{Unemployment}}  & \multicolumn{2}{c}{\textbf{Consumption}}\\
& \multicolumn{1}{c}{\textbf{Mean}} & \multicolumn{1}{c}{\textbf{Std}} & \multicolumn{1}{c}{\textbf{Mean}} & \multicolumn{1}{c|}{\textbf{Std}} & \multicolumn{1}{c}{\textbf{Mean}} & \multicolumn{1}{c}{\textbf{Std}} & \multicolumn{1}{c}{\textbf{Mean}} & \multicolumn{1}{c}{\textbf{Std}} & \multicolumn{1}{c}{\textbf{Mean}} & \multicolumn{1}{c}{\textbf{Std}} \\
\hline
1 & -0.009 & 1 & -0.004 & 0.997 & -0.006 & 0.991 & 0.004 & 1.006 & 0.017 & 1.002 \\
2 & -0.014 & 1.002 & -0.008 & 0.998 & -0.013 & 0.992 & 0.005 & 1.004 & 0.013 & 1.001 \\
3 & -0.017 & 1.003 & -0.008 & 0.999 & -0.015 & 0.992 & 0.007 & 1.003 & 0.014 & 1.001 \\
4 & -0.018 & 1.003 & -0.009 & 0.999 & -0.017 & 0.992 & 0.008 & 1.004 & 0.014 & 1 \\
5 & -0.02 & 1.003 & -0.01 & 0.999 & -0.018 & 0.992 & 0.008 & 1.003 & 0.016 & 1 \\
6 & -0.021 & 1.003 & -0.009 & 0.999 & -0.019 & 0.993 & 0.007 & 1.005 & 0.015 & 1 \\
7 & -0.022 & 1.002 & -0.009 & 0.998 & -0.02 & 0.992 & 0.007 & 1.006 & 0.016 & 1 \\
8 & -0.022 & 1.002 & -0.009 & 0.998 & -0.02 & 0.993 & 0.006 & 1.004 & 0.016 & 1 \\
9 & -0.023 & 1.002 & -0.009 & 0.997 & -0.021 & 0.993 & 0.006 & 1 & 0.017 & 1 \\
10 & -0.024 & 1.002 & -0.009 & 0.997 & -0.021 & 0.993 & 0.006 & 0.999 & 0.017 & 1 \\
11 & -0.024 & 1.001 & -0.008 & 0.996 & -0.022 & 0.993 & 0.006 & 0.998 & 0.018 & 1.001 \\
12 & -0.024 & 1.001 & -0.009 & 0.995 & -0.023 & 0.993 & 0.005 & 0.996 & 0.017 & 1.001 \\
\end{tabular}
\end{table}

\begin{table}[h]
\centering
\begin{tabular}{c|cccccccc}
\multicolumn{1}{c|}{} & \multicolumn{8}{c}{\textbf{4 Variables}} \\
\multicolumn{1}{c|}{\textbf{Forecasting horizon}} & \multicolumn{2}{c}{\textbf{CPI}} & \multicolumn{2}{c}{\textbf{Unemployment}} & \multicolumn{2}{c}{\textbf{Consumption}} & \multicolumn{2}{c}{\textbf{Loans}} \\
& \multicolumn{1}{c}{\textbf{Mean}} & \multicolumn{1}{c}{\textbf{Std}} & \multicolumn{1}{c}{\textbf{Mean}} & \multicolumn{1}{c}{\textbf{Std}} & \multicolumn{1}{c}{\textbf{Mean}} & \multicolumn{1}{c}{\textbf{Std}} & \multicolumn{1}{c}{\textbf{Mean}} & \multicolumn{1}{c}{\textbf{Std}} \\
\hline
1 & 0.005 & 0.99 & -0.001 & 0.989 & -0.006 & 0.997 & 0 & 0.993 \\
2 & 0 & 0.99 & 0.002 & 0.995 & -0.006 & 0.998 & 0 & 0.999 \\
3 & 0 & 0.99 & 0.002 & 0.992 & -0.007 & 0.998 & 0.002 & 1.001 \\
4 & 0.002 & 0.991 & 0.002 & 1.005 & -0.007 & 0.997 & 0.001 & 0.996 \\
5 & 0.004 & 0.991 & 0.003 & 1.005 & -0.008 & 0.997 & -0.001 & 0.995 \\
6 & 0.004 & 0.991 & 0.004 & 1.002 & -0.008 & 0.997 & -0.001 & 0.995 \\
7 & 0.005 & 0.991 & 0.005 & 1.002 & -0.009 & 0.997 & -0.001 & 0.993 \\
8 & 0.004 & 0.991 & 0.005 & 1.001 & -0.008 & 0.998 & -0.002 & 0.995 \\
9 & 0.004 & 0.992 & 0.006 & 1.002 & -0.009 & 0.998 & -0.002 & 0.993 \\
10 & 0.004 & 0.992 & 0.006 & 0.999 & -0.008 & 0.998 & -0.003 & 0.993 \\
11 & 0.004 & 0.992 & 0.007 & 0.997 & -0.009 & 0.998 & -0.002 & 0.991 \\
12 & 0.004 & 0.992 & 0.006 & 0.996 & -0.009 & 0.999 & -0.003 & 0.991 \\
\end{tabular}
\end{table}
\restoregeometry

\begin{table}[h]
\centering
\captionsetup{font=bf, justification=centering}
\caption{\textbf{Verification of distributions of products of standardized forecast errors for unconditional forecast in ABM}}
\begin{tabular}{c|cccc|cccccc}
\multicolumn{1}{c|}{} & \multicolumn{4}{c|}{\textbf{2 variables}} & \multicolumn{6}{c}{\textbf{3 variables}} \\
\multicolumn{1}{c|}{\textbf{Forecasting horizon}} & \multicolumn{2}{c}{\textbf{CPI}} & \multicolumn{2}{c|}{\textbf{Unemployment}} & \multicolumn{2}{c}{\textbf{CPI}} & \multicolumn{2}{c}{\textbf{Unemployment}}  & \multicolumn{2}{c}{\textbf{Consumption}}\\
& \multicolumn{1}{c}{\textbf{Mean}} & \multicolumn{1}{c}{\textbf{Std}} & \multicolumn{1}{c}{\textbf{Mean}} & \multicolumn{1}{c|}{\textbf{Std}} & \multicolumn{1}{c}{\textbf{Mean}} & \multicolumn{1}{c}{\textbf{Std}} & \multicolumn{1}{c}{\textbf{Mean}} & \multicolumn{1}{c}{\textbf{Std}} & \multicolumn{1}{c}{\textbf{Mean}} & \multicolumn{1}{c}{\textbf{Std}} \\
\hline
1 & 0.005 & 1.028 & 0 & 0.975 & 0.015 & 1.013 & -0.011 & 0.998 & -0.006 & 1.017 \\
2 & 0.008 & 1.018 & 0.003 & 0.971 & 0.007 & 1.002 & -0.005 & 0.983 & 0 & 1.011 \\
3 & 0.007 & 1.016 & 0.002 & 0.971 & 0.004 & 0.997 & -0.004 & 0.979 & -0.001 & 1.011 \\
4 & 0.004 & 1.018 & 0.003 & 0.974 & 0.003 & 0.997 & -0.003 & 0.979 & -0.001 & 1.005 \\
5 & 0.004 & 1.017 & 0.001 & 0.975 & 0.001 & 0.997 & -0.004 & 0.976 & -0.004 & 1.006 \\
6 & 0.003 & 1.018 & 0.001 & 0.973 & -0.001 & 0.998 & -0.005 & 0.974 & -0.005 & 1.004 \\
7 & 0.002 & 1.018 & 0 & 0.976 & -0.003 & 1 & -0.005 & 0.974 & -0.005 & 1.003 \\
8 & 0.002 & 1.019 & 0 & 0.979 & -0.004 & 1.002 & -0.005 & 0.976 & -0.006 & 1.004 \\
9 & 0.002 & 1.019 & -0.001 & 0.982 & -0.006 & 1.003 & -0.006 & 0.979 & -0.007 & 1.007 \\
10 & 0.002 & 1.02 & -0.004 & 0.982 & -0.008 & 1.004 & -0.009 & 0.978 & -0.006 & 1.009 \\
11 & 0.002 & 1.017 & -0.005 & 0.983 & -0.009 & 1.005 & -0.01 & 0.979 & -0.009 & 1.01 \\
12 & 0.002 & 1.016 & -0.005 & 0.988 & -0.009 & 1.004 & -0.011 & 0.981 & -0.008 & 1.009 \\
\end{tabular}
\end{table}

\begin{table}[h]
\centering
\begin{tabular}{c|cccccccc}
\multicolumn{1}{c|}{} & \multicolumn{8}{c}{\textbf{4 Variables}} \\
\multicolumn{1}{c|}{\textbf{Forecasting horizon}} & \multicolumn{2}{c}{\textbf{CPI}} & \multicolumn{2}{c}{\textbf{Unemployment}} & \multicolumn{2}{c}{\textbf{Consumption}} & \multicolumn{2}{c}{\textbf{Loans}} \\
\multicolumn{1}{c|}{} & \multicolumn{1}{c}{\textbf{Mean}} & \multicolumn{1}{c}{\textbf{Std}} & \multicolumn{1}{c}{\textbf{Mean}} & \multicolumn{1}{c}{\textbf{Std}} & \multicolumn{1}{c}{\textbf{Mean}} & \multicolumn{1}{c}{\textbf{Std}} & \multicolumn{1}{c}{\textbf{Mean}} & \multicolumn{1}{c}{\textbf{Std}} \\
\hline
1 & 0.02 & 1.033 & 0 & 1.015 & -0.014 & 1.039 & 0.014 & 1.014 \\
2 & 0.016 & 1.013 & 0.004 & 1.012 & -0.005 & 1.025 & 0.001 & 1.013 \\
3 & 0.009 & 1.007 & 0.001 & 1.009 & -0.005 & 1.022 & 0 & 1.007 \\
4 & 0.008 & 1.005 & 0 & 1 & -0.003 & 1.012 & 0 & 1 \\
5 & 0.007 & 1.005 & -0.002 & 1.001 & -0.005 & 1.012 & 0 & 1.002 \\
6 & 0.005 & 1.006 & -0.004 & 0.995 & -0.007 & 1.008 & 0.001 & 1.008 \\
7 & 0.002 & 1.009 & -0.007 & 0.996 & -0.008 & 1.008 & -0.001 & 1.011 \\
8 & 0 & 1.011 & -0.007 & 0.994 & -0.008 & 1.007 & -0.001 & 1.014 \\
9 & -0.001 & 1.014 & -0.01 & 1 & -0.01 & 1.011 & -0.003 & 1.014 \\
10 & -0.004 & 1.015 & -0.013 & 0.996 & -0.009 & 1.013 & -0.003 & 1.017 \\
11 & -0.005 & 1.014 & -0.015 & 0.996 & -0.013 & 1.013 & -0.006 & 1.02 \\
12 & -0.006 & 1.014 & -0.018 & 0.999 & -0.013 & 1.014 & -0.007 & 1.026 \\
\end{tabular}
\end{table}

\clearpage
\newpage

\begin{table}[h]
\centering
\captionsetup{font=bf, justification=centering}
\caption{\textbf{Comparison of unconditional forecasting properties of VAR model and neural network approximating ABM (MSFE is the ratio of the mean squared errors of the VAR and the neural network; LPS is the difference in the mean log predictive scores of the VAR and the neural network)}}
\begin{adjustbox}{valign=t, max width=1.05\textwidth}
\begin{tabular}{c|cccc|cccccc}
\multicolumn{1}{c|}{} & \multicolumn{4}{c|}{\textbf{2 variables}} & \multicolumn{6}{c}{\textbf{3 variables}} \\
\multicolumn{1}{c|}{\textbf{Forecasting}} & \multicolumn{2}{c}{\textbf{CPI}} & \multicolumn{2}{c|}{\textbf{Unemployment}} & \multicolumn{2}{c}{\textbf{CPI}} & \multicolumn{2}{c}{\textbf{Unemployment}}  & \multicolumn{2}{c}{\textbf{Consumption}}\\
\multicolumn{1}{c|}{\textbf{horizon}}& \multicolumn{1}{c}{\textbf{Mean}} & \multicolumn{1}{c}{\textbf{Std}} & \multicolumn{1}{c}{\textbf{Mean}} & \multicolumn{1}{c|}{\textbf{Std}} & \multicolumn{1}{c}{\textbf{Mean}} & \multicolumn{1}{c}{\textbf{Std}} & \multicolumn{1}{c}{\textbf{Mean}} & \multicolumn{1}{c}{\textbf{Std}} & \multicolumn{1}{c}{\textbf{Mean}} & \multicolumn{1}{c}{\textbf{Std}} \\
\hline
1 & 1.05*** & -0.17*** & 1.09*** & -0.58*** & 1.06*** & -0.11*** & 1.11*** & -0.18*** & 1.02** & -0.11** \\
2 & 1.08*** & -0.15*** & 1.11*** & -0.55*** & 1.08*** & -0.16*** & 1.13*** & -0.17*** & 1.03*** & -0.08*** \\
3 & 1.09*** & -0.18*** & 1.12*** & -0.55*** & 1.1*** & -0.2*** & 1.14*** & -0.18*** & 1.05*** & -0.04*** \\
4 & 1.1*** & -0.21*** & 1.14*** & -0.57*** & 1.11*** & -0.23*** & 1.16*** & -0.19*** & 1.06*** & -0.05*** \\
5 & 1.11*** & -0.25*** & 1.16*** & -0.59*** & 1.11*** & -0.25*** & 1.17*** & -0.2*** & 1.08*** & -0.06*** \\
6 & 1.11*** & -0.27*** & 1.17*** & -0.6*** & 1.11*** & -0.27*** & 1.18*** & -0.2*** & 1.09*** & -0.06*** \\
7 & 1.11*** & -0.29*** & 1.18*** & -0.62*** & 1.12*** & -0.28*** & 1.19*** & -0.21*** & 1.1*** & -0.07*** \\
8 & 1.11*** & -0.31*** & 1.19*** & -0.64*** & 1.12*** & -0.28*** & 1.2*** & -0.22*** & 1.1*** & -0.07*** \\
9 & 1.12*** & -0.33*** & 1.2*** & -0.66*** & 1.12*** & -0.28*** & 1.2*** & -0.22*** & 1.11*** & -0.07*** \\
10 & 1.12*** & -0.34*** & 1.2*** & -0.68*** & 1.13*** & -0.28*** & 1.2*** & -0.23*** & 1.11*** & -0.07*** \\
11 & 1.12*** & -0.35*** & 1.2*** & -0.7*** & 1.13*** & -0.28*** & 1.21*** & -0.24*** & 1.11*** & -0.08*** \\
12 & 1.13*** & -0.37*** & 1.21*** & -0.7*** & 1.13*** & -0.28*** & 1.21*** & -0.24*** & 1.12*** & -0.08*** \\
\end{tabular}
\end{adjustbox}
\end{table}

\begin{table}[h]
\captionsetup{labelfont={normalfont}}
\centering
\begin{tabular}{c|cccccccc}
\multicolumn{1}{c|}{} & \multicolumn{8}{c}{\textbf{4 Variables}} \\
\multicolumn{1}{c|}{\textbf{Forecasting horizon}} & \multicolumn{2}{c}{\textbf{CPI}} & \multicolumn{2}{c}{\textbf{Unemployment}} & \multicolumn{2}{c}{\textbf{Consumption}} & \multicolumn{2}{c}{\textbf{Loans}} \\
\multicolumn{1}{c|}{} & \multicolumn{1}{c}{\textbf{Mean}} & \multicolumn{1}{c}{\textbf{Std}} & \multicolumn{1}{c}{\textbf{Mean}} & \multicolumn{1}{c}{\textbf{Std}} & \multicolumn{1}{c}{\textbf{Mean}} & \multicolumn{1}{c}{\textbf{Std}} & \multicolumn{1}{c}{\textbf{Mean}} & \multicolumn{1}{c}{\textbf{Std}} \\
\hline
1 & 1.07*** & -0.11*** & 1.09*** & -0.33*** & 0.980 & -0.09** & 1.35*** & -0.96*** \\
2 & 1.09*** & -0.16*** & 1.1*** & -0.3*** & 1.06*** & -0.07*** & 1.24*** & -0.93*** \\
3 & 1.12*** & -0.21*** & 1.12*** & -0.29*** & 1.08*** & -0.06*** & 1.25*** & -1.02*** \\
4 & 1.13*** & -0.25*** & 1.14*** & -0.27*** & 1.11*** & -0.07*** & 1.28*** & -1.12*** \\
5 & 1.14*** & -0.27*** & 1.15*** & -0.26*** & 1.12*** & -0.07*** & 1.32*** & -1.2*** \\
6 & 1.15*** & -0.28*** & 1.16*** & -0.26*** & 1.14*** & -0.08*** & 1.35*** & -1.28*** \\
7 & 1.16*** & -0.28*** & 1.17*** & -0.25*** & 1.16*** & -0.08*** & 1.39*** & -1.35*** \\
8 & 1.17*** & -0.29*** & 1.18*** & -0.25*** & 1.19*** & -0.08*** & 1.42*** & -1.41*** \\
9 & 1.18*** & -0.29*** & 1.18*** & -0.25*** & 1.22*** & -0.09*** & 1.45*** & -1.47*** \\
10 & 1.19*** & -0.29*** & 1.19*** & -0.26*** & 1.27*** & -0.09*** & 1.48*** & -1.52*** \\
11 & 1.2*** & -0.29*** & 1.19*** & -0.26*** & 1.28*** & -0.1*** & 1.5*** & -1.57*** \\
12 & 1.21*** & -0.29*** & 1.19*** & -0.27*** & 1.29*** & -0.1*** & 1.53*** & -1.62*** \\
\end{tabular}

\caption*{Significance level: *(10\%), **(5\%), ***(1\%)}
\end{table}

\FloatBarrier
\newgeometry{left=2.5cm, right=2cm, top=0cm, bottom=2.5cm}
\begin{figure}[h]
  \centering
  \textbf{Figure 5. Neural network forecasts for ABM with 4 variables}
  \scalebox{1}{\includegraphics{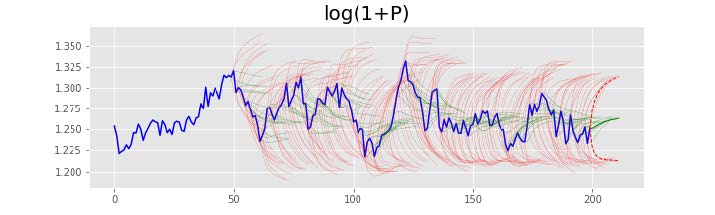}}
  \scalebox{1}{\includegraphics{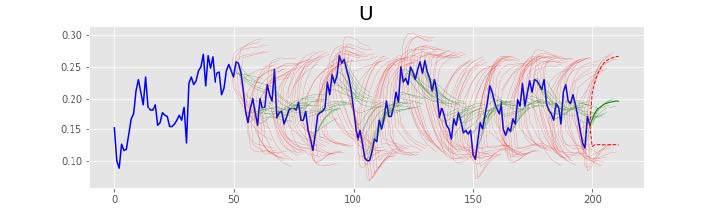}}
  \scalebox{1}{\includegraphics{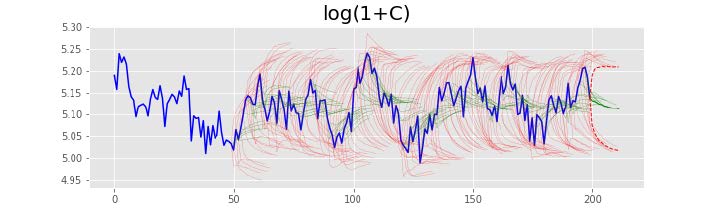}}
  \scalebox{1}{\includegraphics{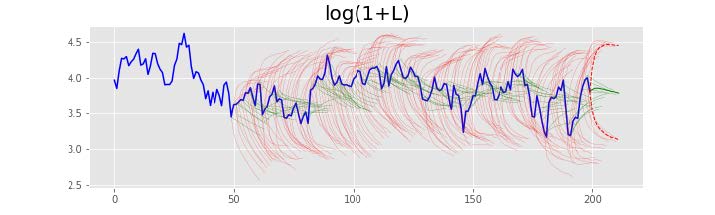}}
\end{figure} 
\restoregeometry
\FloatBarrier

\begin{table}[h]
\centering
\captionsetup{font=bf, justification=centering}
\caption{Comparison of conditional forecasting properties of VAR model and neural network approximating ABM (MSFE is the ratio of the mean squared errors of the VAR and the neural network; LPS is the difference in the mean log predictive scores of the VAR and the neural network)}
\begin{tabular}{c|cc}

\textbf{Forecasting horizon} & \textbf{MSFE} & \textbf{LPS} \\
\hline
1 & 1.231*** & -0.242*** \\
2 & 1.276*** & -0.235*** \\
3 & 1.318*** & -0.199*** \\
4 & 1.349*** & -0.216*** \\
5 & 1.38*** & -0.236*** \\
6 & 1.379*** & -0.234*** \\
7 & 1.389*** & -0.24*** \\
8 & 1.394*** & -0.247*** \\
9 & 1.392*** & -0.254*** \\
10 & 1.405*** & -0.26*** \\
11 & 1.374*** & -0.276*** \\
12 & 1.284*** & -0.215*** \\
\end{tabular}
\label{tab:forecasting}
\caption*{\textnormal{Significance level: *(10\%), **(5\%), ***(1\%)}}
\end{table}

\FloatBarrier

\begin{figure}[h]
  \centering
  \textbf{Figure 6. Neural network conditional forecasts for ABM}
  \scalebox{1}{\includegraphics{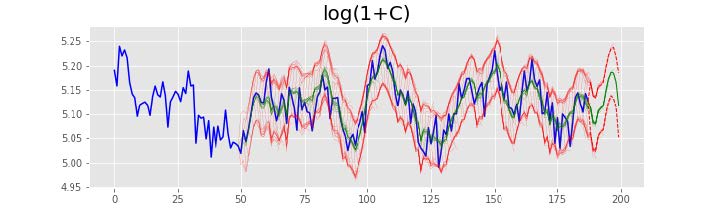}}
\end{figure} 

\newgeometry{left=2.5cm, right=2cm, top=2.5cm, bottom=2.5cm}
\begin{figure}[h]
  \centering
  \textbf{Figure 7. Distributions of standardized forecast errors for conditional forecast from ABM}
  \scalebox{1}{\includegraphics{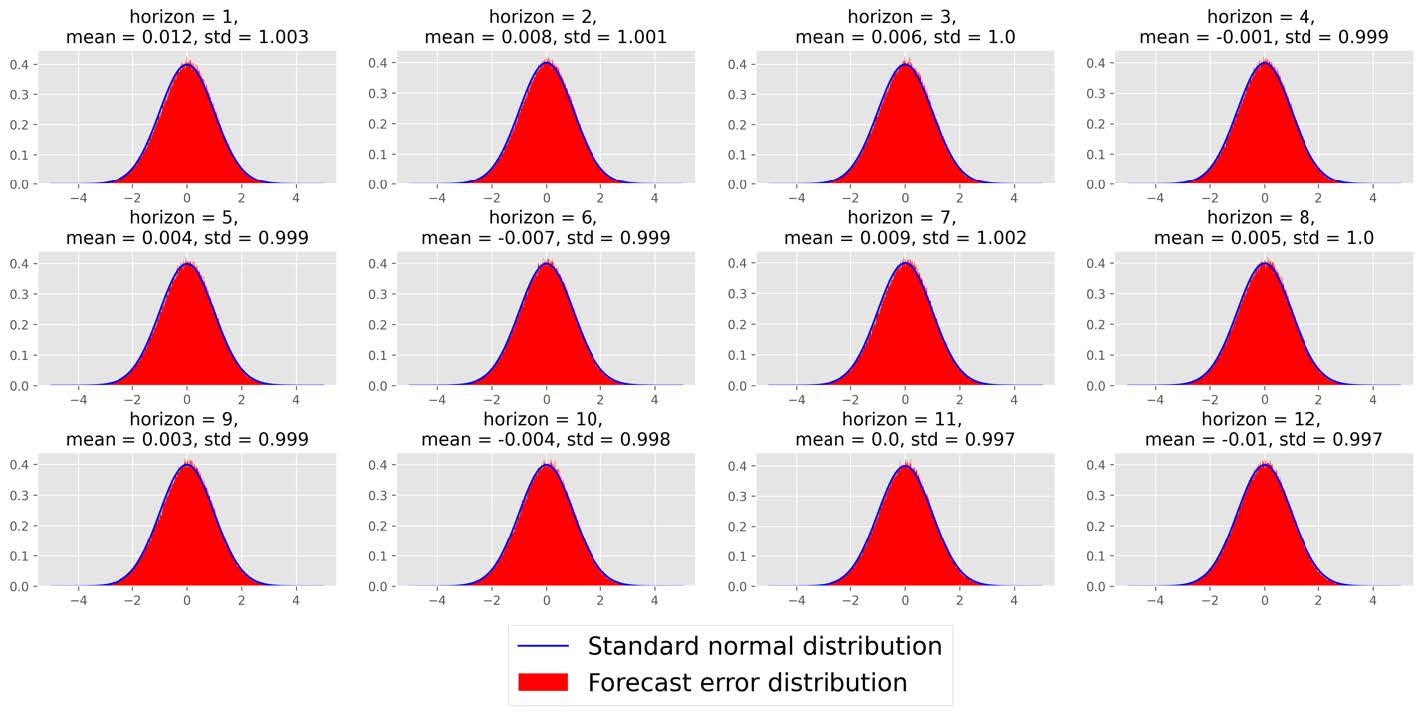}}
\end{figure} 

\begin{figure}[h]
  \centering
  \textbf{Figure 8. Distributions of products of  standardized  forecast errors for conditional forecast from ABM}
  
  \scalebox{1}{\includegraphics{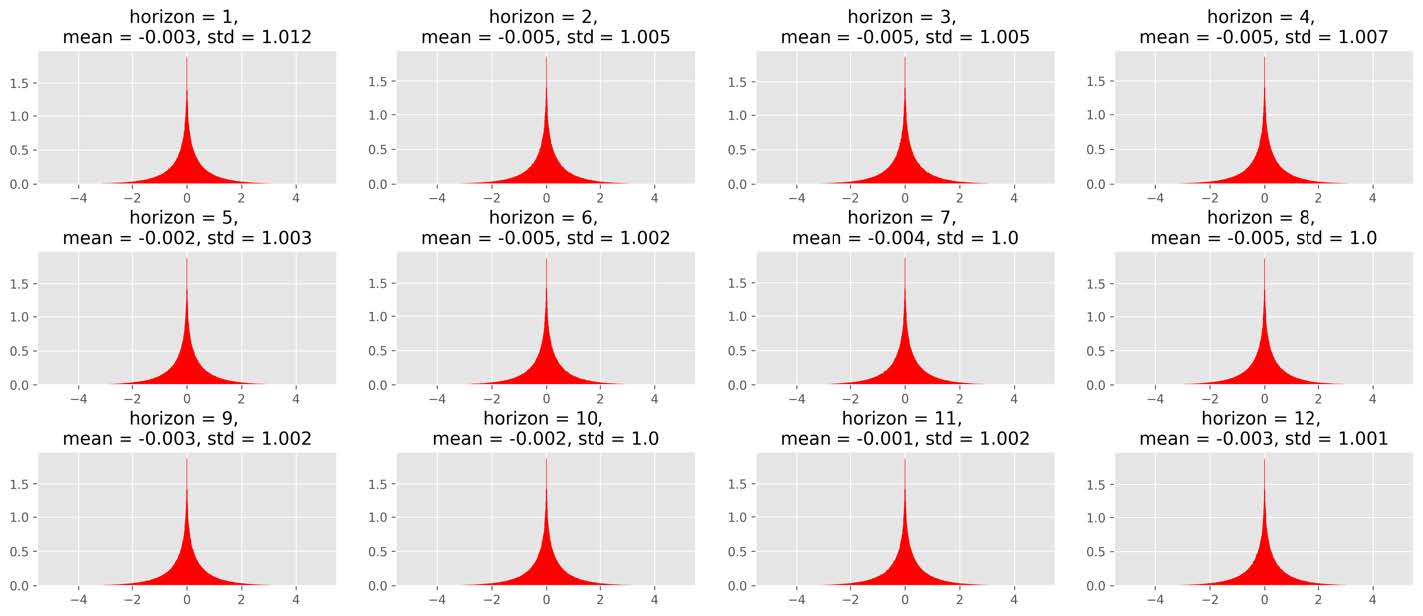}}
\end{figure} 
\FloatBarrier
\restoregeometry
\newpage
\section* {Appendix B. Models}
      
\subsection* {B.1. Bayesian AR(1) model}
\textit{Prior distribution:}
\[\sigma^2 \sim IG(3;1), \quad \rho \sim N(0; \sigma), \quad if \quad 0 \leq \rho<1\]
where $\sigma$ is the standard deviation of shocks, $\rho$ is the autoregression coefficient, $IG(a,b)$ is the inverse gamma distribution with parameters $a$ and $b$, and $N(m,s)$ is the normal distribution with mean m and standard deviation $s$.

\noindent\textit{Data generation process:}

\[y_0 \sim N(0; \frac{\sigma}{\sqrt{1-\rho^2}})\]
\[y_t=\rho y_{t-1}+e_t,		 \quad t=1,...,T \]
\[e_t \sim N(0;\sigma) ,	\quad	t=1,...,T\]
where $y_t$ is the value of the time series at time $t$ and $e_t$ is the random shock at time $t$.

\subsection*      {B.2. ABM}
      The model consists of C-firms, consumers (workers), a bank, and the government. The C-firms produce goods using linear technology:
\[y_{i,t}=\alpha L_{i,t} \]
where $\alpha$ is the labour productivity, $y_{i,t}$ is the volume of goods produced by firm $i$ in period $t$, and $L_{i,t}$ is the amount of labour used for production. The firm takes several steps before producing goods.

      First, the firms set their prices and choose their desired output level based on rules similar to those proposed by \cite{8}:
\[P_{i,t}^{lower}=\frac{w}{\alpha} + \frac{r_{i-1,t} B_{i,t-1}} {max(y_{i,t-1},\alpha)} \]

\[y_{i,t}^*=max(y_{i,t-1}+\rho_{i,t} (I(P_{i,t-1}>P_{t-1} )I(\Delta_{i,t-1}>0)-I(P_{i,t-1}<P_{t-1} )I(\Delta_{i,t-1} \leq 0)) y_{i,t-1}, \alpha) \]

\[P_{i,t}=max(P_{i,t-1}+\eta_{i,t} (I(P_{i,t-1}<P_{t-1} )I(\Delta_{i,t-1}>0)-I(P_{i,t-1}>P_{t-1} )I(\Delta_{i,t-1} \leq 0))P_{i,t-1},P_{i,t}^{lower} ) \]
where $P_{i,t}^{lower}$ is the minimum price of firm $i$ in period $t$, $w$ is the wage of workers, $r_{i-1,t}$ is the average interest rate on the loans of firm $i$ in period $t-1$, $B_{i,t-1}$ is the volume of loans of firm $i$ in period $t-1$, $y_{i,t}^*$ is the desired output of firm $i$ in period $t$, $P_{i,t}$ is the price of the goods of firm $i$ in period $t$, $\rho_{i,t}$ and $\eta_{i,t}$ are random variables from the uniform distributions with zero lower bounds and upper bounds $\rho$ and $\eta$, respectively, $P_{t-1}$ is the arithmetic average price of goods in period $t-1$, $\Delta_{i,t-1}$ is the excess demand for the products of firm $i$ in period $t-1$, and $I$ is an indicator equal to one if the condition is satisfied and equal to zero if it is not.

      When the prices and the desired level of output have been formed, the firms determine the amount of money needed to finance production. This amount is determined by the wages that must be paid to the workers involved in production. The number of workers the firm needs to produce the desired level of output is calculated by the formula:
      \[ L_{i,t}^*= \frac{y_{i,t}^*}{\alpha} \]
If the firm understands that its own liquidity $(M_{i,t-1})$ is not enough, it approaches the bank for the necessary financing $(wL_{i,t}^*-M_{i,t-1})$.

      The bank decides on the maximum amount of credit to be granted to firm $i$ on the basis of the fulfilment of the capital requirements and the loans already granted. The total amount of financing provided to firm $i$ is:
      
\[B_{i,t}^{new}=min(max(wL_{i,t}^*-M_{i,t-1}, 0), \frac {E_{t-1}}{\zeta N_c}-B_{i,t-1}) \]
where $E_{t-1}$ is the bank's capital in period $t-1$, $\zeta$ is a constant responsible for the capital adequacy ratio,\footnote{Not necessarily equal to it.} and $N_c$ is the number of C-firms. The interest rate at which the new loan is granted is set taking into account the company's desired debt-to-asset ratio $(\lambda_{i,t})$:

\[r_{i,t}^{new}=r(1+\mu_{i,t} \lambda_{i,t}^{0.5}) \]
\[\lambda_{i,t}=\frac{max(wL_{i,t}^*-M_{i,t-1}, 0) +B_{i,t-1}}{max(wL_{i,t}^*-M_{i,t-1}, 0) +B_{i,t-1}+E_{i,t-1}^c+10^{-8}} \]
where $r_{i,t}^{new}$ is the interest rate on the loan received by firm $i$ in period $t$, $r$ is the risk-free interest rate, and $\mu_{i,t}$ is a random variable from a uniform distribution with a zero lower bound and upper bound $\mu$. The new volume of loans $B_{i,t}^*$ and the average interest rate $r_{i,t}$ for firm $i$ are thus:
\[B_{i,t}^*=B_{i,t}^{new}+B_{i,t-1} \]
\[r_{i,t}=\frac {B_{i,t}^{new}}{B_{i,t}^*} r_{i,t}^{new} + \frac {B_{i,t-1}}{B_{i,t}^*} r_{i-1,t} \]

      The final step before production is the hiring of labour. The firms post vacancies to find workers knowing the level of funds available and the desired output. Each firm posts new vacancies if

\[V_{i,t}=round(min(L_{i,t}^*-L_{i,t-1}, \frac {M_{i,t-1}+B_{i,t}^{new}}{w} ))>0 \]
and fires employees if $V_{i,t}<0$. The labour market operates in $Z_e$ rounds, and the workers are sorted randomly. In each round, the workers approach their employers, and if the firm has more labour than it needs, it fires those who arrive earliest. The unemployed approach randomly selected firms and get jobs if there are unfilled vacancies.

      After the labour market stage, the firms produce goods, pay wages, and sell their goods. In the goods market, a search mechanism similar to that described for the labour market operates. The consumers form their desired consumption budgets $(CB_{j,t}^*)$, which consist of estimates of permanent income $(PI_{j,t})$ and current wealth $(W_{j,t})$:
      
\[PI_{j,t}= \Sigma PI_{j,t-1}+(1-\Sigma) I_{j,t} \]
\[CB_{j,t}^*= \psi PI_{j,t} + \chi W_{j,t} \]
where $I_{j,t}$ is the income of consumer $j$ in time period $t$, equal to the sum of dividends minus the recapitalization of bankrupt firms (see below) and wages after taxes, if the consumer is employed, and unemployment benefits $zw$, if the consumer is unemployed, $\Sigma$ is the smoothing parameter in estimating permanent income, and $\psi$ and $\chi$ are the shares of permanent income and wealth, which are used to form the desired budget for consumption. If the desired consumption budget is less than current welfare, it is the consumption budget, otherwise the consumption budget is half of current welfare. The consumers randomly select $Z_c$ firms and visit them in $Z_c$ rounds with given consumption budgets. In the first round, the consumer goes to the firm with the lowest price among those selected and buys goods from it. If the firm does not have enough goods to meet the demand of the consumer, the consumer buys whatever is available and saves the remaining budget for the next round.

      Each period, the firms return fixed share $\theta$ of debt to the banks. The firms with positive profits pay dividends to households (all equally) equal to the minimum of share $\tau$ of profit and the current liquidity held by the firm. The firms that have negative equity after all the previous steps are bankrupt (the bank loses the difference between the loans and the firms' liquidity). All their employees become unemployed. These firms are replaced by new firms, which enter the market with zero debt and with equity and liquidity equal to the average equity of the non-bankrupt firms. The creation of firms is financed by households in equal proportions. If the total wealth of households is less than the amount of money needed to create new firms, the size of new firms is proportionally reduced. Also, if a household does not have enough wealth to finance all firms in the recapitalization step, it receives a non-repayable transfer from the government in the amount of the missing funds.

      The firms must set values for variables such as price, previous output, and demand when they enter the market. We consider two variants of how this is done, assuming in a certain sense a uniform prior distribution for the two different models. In the first of them, the firms set the variables based on the values of the predecessor firm; in the second, the firms set the values on the basis of market averages.

      At this point the period ends, and all aggregate variables are calculated and added to the observed dataset as the variables of period $t$.

\begin{table}[h]
\centering
\caption*{\textbf{Table B1. Parameter values and prior distributions}}
\begin{tabular}{c|c|c|c}

\multirow{2}{*}{\textbf{Parameter}} & \multirow{2}{*}{\textbf{Description}} & \multirow{2}{*}{\textbf{Value}} & \multirow{1}{*}{\textbf{Prior}}  \\
\multirow{2}{*}{} & \multirow{2}{*}{} & \multirow{2}{*}{} & \multirow{1}{*}{\textbf{distribution}} \\
\hline
$N_w$ & Number of households & 500 & - \\
$N_c$ & Number of C-firms & 50 & - \\
$Z_c$ & Number of rounds in goods market & 2 & - \\
$Z_e$ & Number of rounds in labour market & 4 & - \\
$\alpha$ & Labour productivity & 0.5 & - \\
$\tau$ & Share of dividends & 0.2 & - \\
$r$ & Risk-free interest rate & 0.01 & - \\
$w$ & Wage & 1 & - \\
$z$ & Unemployment benefit to wage ratio & 0.5 & - \\
$\Sigma$ & Smoothing parameter in estimating permanent income & - & $Uniform(0,1)$ \\
$\psi$ & Share of permanent income in consumption budget & - & $Uniform(0.5,1)$ \\
$\chi$ & Share of welfare in consumption budget & - & $Uniform(0,1)$ \\
$\rho$ & Upper bound on change in output & - & $Uniform(0,0.5)$ \\
$\eta$ & Upper bound on change in price & - & $Uniform(0,0.5)$ \\
$\mu$ & Upper bound on risk premium & - & $Uniform(0,2)$ \\
$\theta$ & Share of loans returned & - & $Uniform(0.01,0.1)$ \\
$t_w$ & Wage tax & - & $Uniform(0,0.3)$ \\
$Newvalues$ & Variables setting mechanism after bankruptcy & - & $Binominal(2,0.5)$ \\

\end{tabular}
\end{table}

\newpage
\section *  {Appendix C. Conditional forecasting for generalized architecture}
      In this appendix, we run additional experiments for a neural network with a generalized architecture, which can be used in cases other than those in which the scenarios are built based on observable variables. As discussed in Section 3.2, one implementation of such an architecture relies on the addition of scenarios at the level of the fully connected layers. We tried various hyperparameters, including those at the level of the fully connected layers, and show results for a model with 5 layers and 100 neurons on each of them. This model is trained for 10 hours (1,000,000 iterations with a batch size of 100) and shows the best results. The task is the same problem as that used for the conditional forecasting. Despite the fact that it does not contain unobservable variables, this task is quite suitable for testing a more general architecture and its properties, since the model trained in Section 4.3 can be used as a benchmark in estimating the quality.

Figure C1 shows predictions on the same data as is used in the main part of the paper. Although the results are similar, they do nevertheless differ. However, as can be seen from a comparison of the results in Table C1 and Table 5, the model with the generalized architecture yields the worst results. The standardized error test (Figures C2 and C3) also shows the presence of a larger bias in the mean than in the case of the architecture described in Section 3.2. Moreover, despite the fact that the mean biases are in several cases comparable in magnitude to those obtained for the unconditional forecast model (Tables 3 and 4), they are systematic in nature and are not related to the drift of the coefficients as we saw from the training process. However, as noted earlier in Section 5, the development of formal statistical procedures or averaging models is necessary for such conclusions to be made more systematic.

All of the above signals that the alternative architecture considered here does not produce forecasts that approximates the characteristics of the posterior distribution. Nevertheless, Figure C1 demonstrates that the results obtained are close to those of Section 4.3, and in some tasks, this approximation quality may be sufficient. However, the decision regarding the sufficiency of the approximation quality should be made for each specific problem separately, based on the goals of the creation of the model and the risks posed by approximation errors.

\begin{table}[h]
\centering
\captionsetup{justification=centering}
\caption*{\textbf{Table C1. Comparison of conditional forecasting properties of VAR model and neural network approximating ABM, generalized architecture (MSFE is the ratio of the mean squared errors of the VAR and the neural network; LPS is the difference in the mean log predictive scores of the VAR and the neural network)}}
\begin{tabular}{c|c|c}
\textbf{Forecasting horizon} & \textbf{MSFE} & \textbf{LPS} \\
\hline
1 & 1.076*** & -0.175** \\
2 & 1.101*** & -0.161*** \\
3 & 1.127*** & -0.119*** \\
4 & 1.16*** & -0.134*** \\
5 & 1.17*** & -0.148*** \\
6 & 1.167*** & -0.143*** \\
7 & 1.172*** & -0.146*** \\
8 & 1.174*** & -0.152*** \\
9 & 1.18*** & -0.158*** \\
10 & 1.192*** & -0.165*** \\
11 & 1.181*** & -0.183*** \\
12 & 1.112*** & -0.121*** \\

\end{tabular}
\caption*{Significance level: *(10\%), **(5\%), ***(1\%)}
\end{table}

\FloatBarrier
\begin{figure}[h]
  \centering
  \vspace{-1cm}
  \textbf{Figure C1. Neural network conditional forecasts for ABM (upper graph is generalized architecture; lower graph is architecture from Section 3.2)}
  \scalebox{1}{\includegraphics{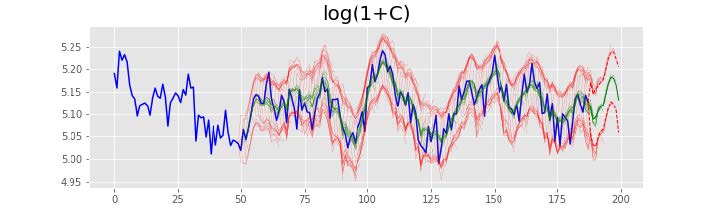}}
  \scalebox{1}{\includegraphics{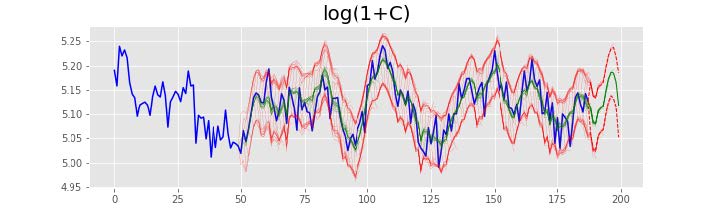}}
\end{figure} 

\begin{figure}[h]
  \centering
  \textbf{Figure C2. Distributions of standardized forecast errors for conditional forecast from ABM, generalized architecture}
  \scalebox{1}{\includegraphics{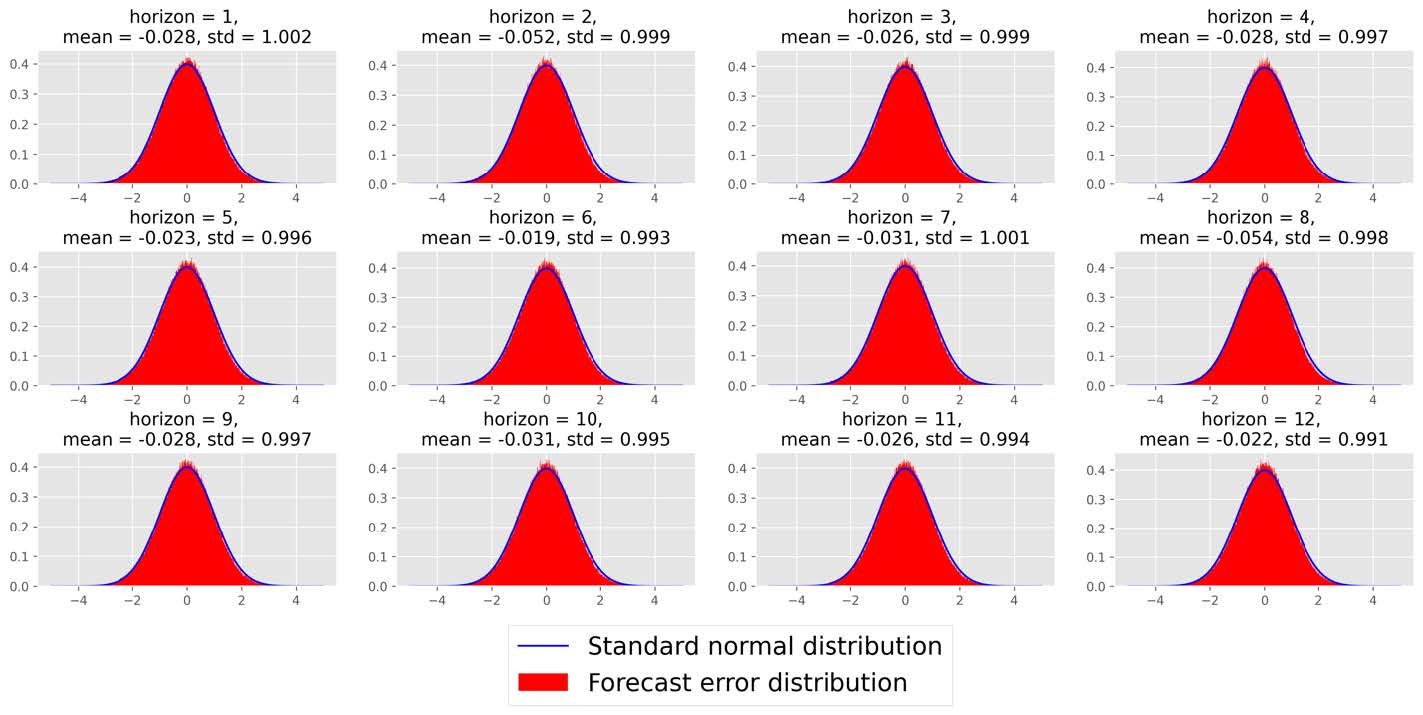}}
\end{figure} 

\newpage
\newgeometry{top = 2.5cm}
\begin{figure}[h]
  \centering
  \textbf{Figure C3. Distributions of products of standardized forecast errors for conditional forecast from ABM, generalized architecture}
  \scalebox{1}{\includegraphics{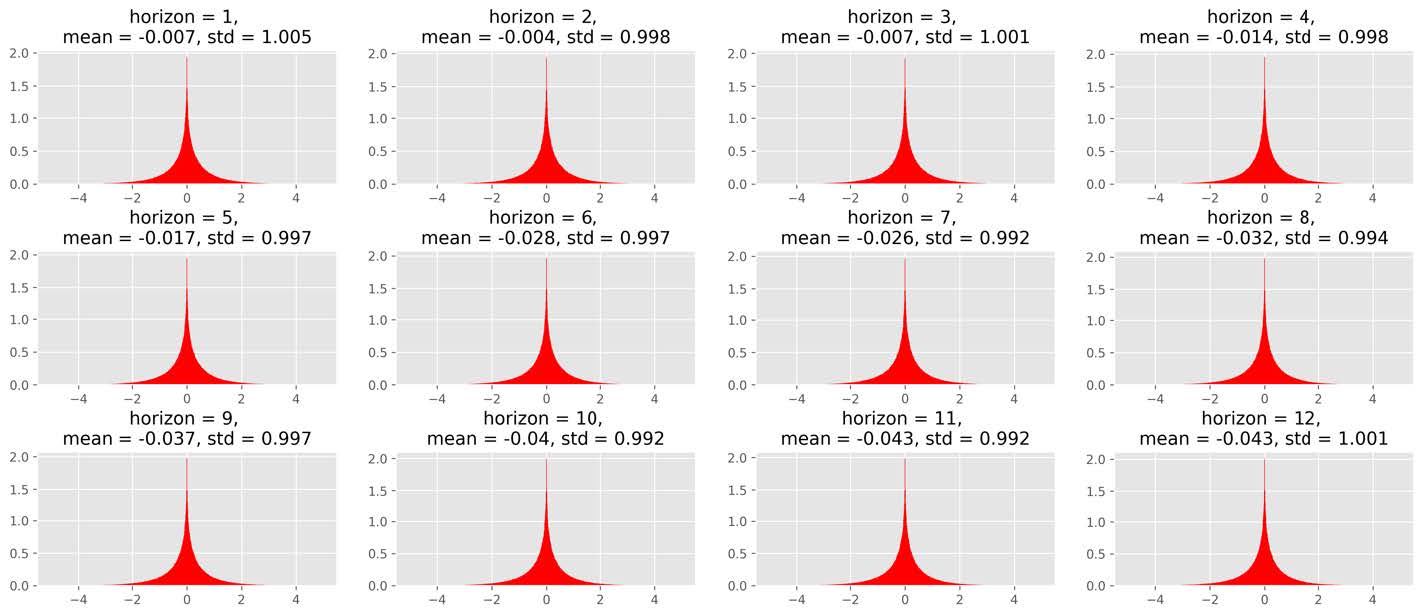}}
\end{figure} 

\end{document}